\documentclass[aip,jcp,reprint,amsmath,amssymb,floatfix,citeautoscript]{revtex4-1}
\usepackage{cancel}
\usepackage{amsmath}
\usepackage{physics}
\usepackage{graphicx}
\usepackage{amssymb}
\usepackage{amsthm}
\usepackage{bm}
\usepackage{dcolumn}
\usepackage{braket}
\usepackage{longtable}
\usepackage{ragged2e}
\usepackage{txfonts}
\usepackage[version=3]{mhchem} 
\usepackage[T1]{fontenc}       
\usepackage{pstricks}
\usepackage{algorithm}         
\usepackage{algpseudocode}     
\usepackage{siunitx}

\usepackage{comment}
\usepackage{footmisc}
\usepackage{ulem}
\usepackage{longtable}
\usepackage{booktabs}

\usepackage[colorlinks=true,allcolors=blue]{hyperref}
\usepackage[capitalise]{cleveref}

\usepackage{threeparttable} 
\usepackage{dcolumn}
\newcolumntype{d}[1]{D{.}{.}{#1}}

\let\red\undefined
\newcommand{\red}[1]{{\textcolor{red}{#1}}}

\let\vr\undefined

\newcommand{\vr}{{\bm{r}}}

\allowdisplaybreaks

\begin{document}

\title{Full-frequency dynamical Bethe-Salpeter equation without frequency and a study of double excitations}

\author{Sylvia J. Bintrim}
\affiliation{Department of Chemistry,
Columbia University, New York, New York 10027, USA}
\author{Timothy C. Berkelbach}
\email{tim.berkelbach@gmail.com}
\affiliation{Department of Chemistry,
Columbia University, New York, New York 10027, USA}
\affiliation{Center for Computational Quantum Physics, Flatiron Institute, New York, New York 10010, USA}

\begin{abstract}
The Bethe-Salpeter equation (BSE) that results from the GW approximation to the
self-energy is a frequency-dependent (nonlinear) eigenvalue problem due to the
dynamically screened Coulomb interaction between electrons and holes. The
computational time required for a numerically exact treatment of this frequency
dependence is $O(N^6)$, where $N$ is the system size.  To avoid the common
static screening approximation, we show that the full-frequency dynamical BSE
can be exactly reformulated as a frequency-independent eigenvalue problem in an
expanded space of single and double excitations. When combined with an iterative
eigensolver and the density fitting approximation to the electron repulsion
integrals, this reformulation yields a dynamical BSE algorithm whose
computational time is $O(N^5)$, which we verify numerically. Furthermore, the reformulation provides
direct access to excited states with dominant double excitation character, which are
completely absent in the spectrum of the statically screened BSE. We study
the $2^1A_\mathrm{g}$ state of butadiene, hexatriene, and octatetraene and find that
GW/BSE overestimates the excitation energy by about 1.5--2~eV and significantly
underestimates the double excitation character.
\end{abstract}

\maketitle

The Bethe-Salpeter equation (BSE) is an exact relation between the two-particle Green's function
and the one-particle self-energy.  Using the GW approximation to the self-energy yields
an approximate solution of the BSE, providing neutral excitation energies and spectra.~\cite{Hedin1965, Strinati1988}
Due to its success in solids,~\cite{hankeManyparticleEffectsOptical1980, strinatiEffectsDynamicalScreening1984, albrechtExcitonicEffectsOptical1998, rohlfingElectronholeExcitationsOptical2000} the BSE has been increasingly applied to molecules~\cite{Tiago2005, Faber2014, Krbel2014, Bruneval2015, Jacquemin2015, Rangel2017, Blase2020}
and compared to common quantum chemistry methods like time-dependent density functional
theory (TDDFT), the algebraic diagrammatic construction (ADC), and coupled-cluster theory.~\cite{Jacquemin2016, Jacquemin2017, Blase2018, Gui2018}
Physically, the BSE modifies the GW energy gaps due to electron-hole 
interactions~\cite{hankeManyparticleEffectsOptical1980, rohlfingElectronholeExcitationsOptical2000}.  Importantly, the electron-hole attraction
in the BSE is screened, which is the primary difference compared to time-dependent
Hartree-Fock (HF) theory.  

In terms of spatial molecular orbitals, the BSE is a frequency-dependent eigenvalue problem
$\bm{\mathcal{A}}(\Omega) \mathbf{X} = \Omega \mathbf{X}$
where 
\begin{subequations}
\label{eq:bse}
\begin{align}
\mathcal{A}_{ia,jb}(\Omega) &= A_{ia,jb} -K^\mathrm{(p)}_{abij}(\Omega) \\
A_{ia,jb} &= (E_a-E_i)\delta_{ij}\delta_{ab}
    +\kappa (ia|jb)-(ab|ij),
\end{align}
\end{subequations}
$E_p$ are GW quasiparticle energies,
$(pq|rs)$ are electron repulsion integrals in $(11|22)$ notation,
and $\kappa=2$ for a singlet excited state and $0$ for a triplet excited state.
Here and throughout, we make the Tamm-Dancoff approximation (TDA), which typically 
introduces negligible error and removes triplet instabilities~\cite{Rangel2017,Blase2018} but can cause significant errors 
when plasmonic effects (collective electronic excitations) are important.\cite{Galli2010}
Furthermore, we use $i,j,k,l$ to index orbitals that are occupied in the mean-field reference; $a,b,c,d$ to index orbitals
that are unoccupied; and $p,q,r,s$ to index general orbitals. For simplicity, we assume real orbitals.

Like the self-energy in the GW approximation, the polarizable part of the direct electron-hole interaction
requires a frequency integration,
\begin{equation}
\begin{split}
\label{eq:freq_int}
&K^{(\mathrm{p})}_{abij}(\Omega) = 
\frac{i}{2\pi} \int d\omega e^{-i\omega 0+} W^{\mathrm{(p)}}_{abij}(\omega) \\
&
\times\Bigg[\frac{1}{\Omega-\omega-(E_b-E_i)+i\eta}
+\frac{1}{\Omega+\omega-(E_a-E_j)+i\eta}\Bigg],
\end{split}
\end{equation}
where
\begin{equation}
W^{(\mathrm{p})}_{abij}(\omega) = \int d\vr_1 d\vr_2 \phi_a(\vr_1) \phi_b(\vr_1) W^{(\mathrm{p})}(\vr_1,\vr_2;\omega)
    \phi_i(\vr_2)\phi_j(\vr_2)
\end{equation}

are matrix elements of the polarizable part of the screened Coulomb interaction
\begin{equation}
\label{eq:Wp}
W^{(\mathrm{p})}(\vr_1,\vr_2;\omega) = \int d\vr \varepsilon^{-1}(\vr_1,\vr;\omega)|\vr-\vr_2|^{-1} - |\vr_1-\vr_2|^{-1}.
\end{equation}

In practice, this frequency integration is typically avoided by making the static screening 
approximation~\cite{Krause2017,Bruneval2016,Liu2020a, Galli2010}, 

\begin{equation}
K_{abij}^{\text{(p)}}(\Omega) \approx W_{abij}^{\text{(p)}}(\omega=0).
\end{equation}
 
Although computationally convenient, the static screening approximation often introduces
errors of about 0.1-0.3~eV~\cite{rohlfingElectronholeExcitationsOptical2000, Ma2009, LoosBlase2020}.
Moreover, the static screening approximation significantly reduces the number of excited states predicted by the BSE;
in particular, as will be discussed more later, it removes double excitations from the BSE 
spectrum~\cite{Romaniello2009, Maitra2011, Authier2020, LoosBlase2020}.
In this work, we show that full-frequency dynamical BSE calculations can be performed by diagonalizing
a frequency-independent Hamiltonian matrix in an expanded space of single and double excitations
(``full-frequency'' means that a model dielectric function or plasmon-pole approximation\cite{rohlfingElectronholeExcitationsOptical2000}
is not used, 
and ``dynamical'' means that the static screening approximation is not used).
This formulation enables the use of iterative eigensolvers and provides access
to BSE eigenvectors with dominant double excitation character, which allows us
to assess the quality of double excitations predicted by the GW/BSE approach.

As the most common example, we will consider the use of the random-phase approximation (RPA)
to calculate the screened Coulomb interaction. 
Like in our previous work~\cite{Bintrim2021}, we consider screening within the TDA to the RPA,
which avoids technical problems associated with positive and negative eigenvalue
pairs in the RPA matrix; the impact of this choice will be assessed with numerical tests, reported below.
Within the TDA, the RPA eigenproblem is $\mathbf{S} \bm{X}^m = \Omega_m \bm{X}^m$,
where $S_{ia,jb} = (\varepsilon_a - \varepsilon_i) \delta_{ab}\delta_{ij} + 2(ia|bj)$.
Using the spectral representation of the RPA polarizability,
the frequency integration~(\ref{eq:freq_int}) can be performed analytically to give
\begin{equation}
\begin{split}
\label{eq:Kd}
&K^{(\mathrm{p})}_{abij}(\Omega) = 2\sum_m^{\Omega_m>0}(ij|\rho_m)(ab|\rho_m) \\
&\hspace{2em}\times \Bigg[\frac{1}{\Omega-(E_b-E_i)-\Omega_m} 
+\frac{1}{\Omega-(E_a-E_j)-\Omega_m}\Bigg],
\end{split}
\end{equation}
where $(pq|\rho_m) = \sum_{ia} X_{ia}^{m} (pq|ia)$ and we have dropped $i\eta$
terms.  The severe disadvantage of the above expression is that it requires an
explicit enumeration of the $O(N^2)$ excitations entering into the
polarizability, which requires diagonalizing the RPA or TDA matrix with $O(N^6)$
cost. Unsurprisingly, given their similar
structure, the same problem plagues full-frequency implementations of the GW
approximation.~\cite{VanSetten2015}

Here we show that the dynamical BSE~(\ref{eq:bse}), with the frequency dependence appearing
as a sum of simple poles as in Eq.~(\ref{eq:Kd}), can be obtained by
downfolding a larger, frequency-independent matrix. This is analogous to what was done in our previous work on the
GW approximation~\cite{Bintrim2021}.  
First, let us assume that the GW eigenvalues $E_p$ 
have been computed.  In that case, it is reasonably straightforward to show that
the self-consistent eigenvalues of $\bm{\mathcal{A}}(\Omega)$ are the same as
those of the frequency-independent matrix
\begin{equation}
\label{eq:hmat}
\bm{\mathcal{H}} =
\left(
\begin{array}{ccc}
\mathbf{A} & -\mathbf{V}^{\text{e}} & -\mathbf{V}^{\text{h}} \\
\mathbf{(V^{\text{h}})^\dagger} & \mathbf{D} & \mathbf{0}
\\
\mathbf{(V^{\text{e}})^\dagger} & \mathbf{0} & \mathbf{D}
\end{array}
\right),
\end{equation}
where
\begin{subequations}
\begin{align}
\mathbf{D} &= [-\bm{E}_{\text{occ}}]\oplus 
    \bm{E}_{\text{vir}}\oplus \mathbf{S} \\
\label{eq:rpa}
V^\text{h}_{ia,ldkc} &= \sqrt{2}(il|kc)\delta_{ad} \\
V^\text{e}_{ia,ldkc} &= \sqrt{2}(kc|ad)\delta_{il}.
\end{align}
\end{subequations}
Note that the orbital energies $\varepsilon_p$ in Eq.~(\ref{eq:rpa}) are the mean-field orbital energies,
e.g.~from DFT, and $\mathbf{S}$ is the direct RPA matrix in the TDA.
The matrix~(\ref{eq:hmat}) can be expressed in a basis of single and double excitations,
similar to configuration interaction or coupled-cluster theories, although the BSE
matrix has two sets of double excitations.
Downfolding the double excitations into the space of single excitations yields the 
frequency-dependent matrix
\begin{align}
\bm{\mathcal{A}}(\omega) &= \mathbf{A}
    -\mathbf{V}^{\text{e}}(\omega \mathbf{I}-\mathbf{D})^{-1}(\mathbf{V}^{\text{h}})^\dagger
    -\mathbf{V}^{\text{h}}(\omega \mathbf{I}-\mathbf{D})^{-1}(\mathbf{V}^{\text{e}})^\dagger,
\end{align}
which can be checked to be identical to the BSE matrix~(\ref{eq:bse}).
The double excitations are therefore responsible for the appearance of screening in the BSE, which
is similar to how they are viewed as allowing for orbital relaxation in quantum chemical theories.

\begin{figure}[b]
  \centering
    \includegraphics[scale=0.85]{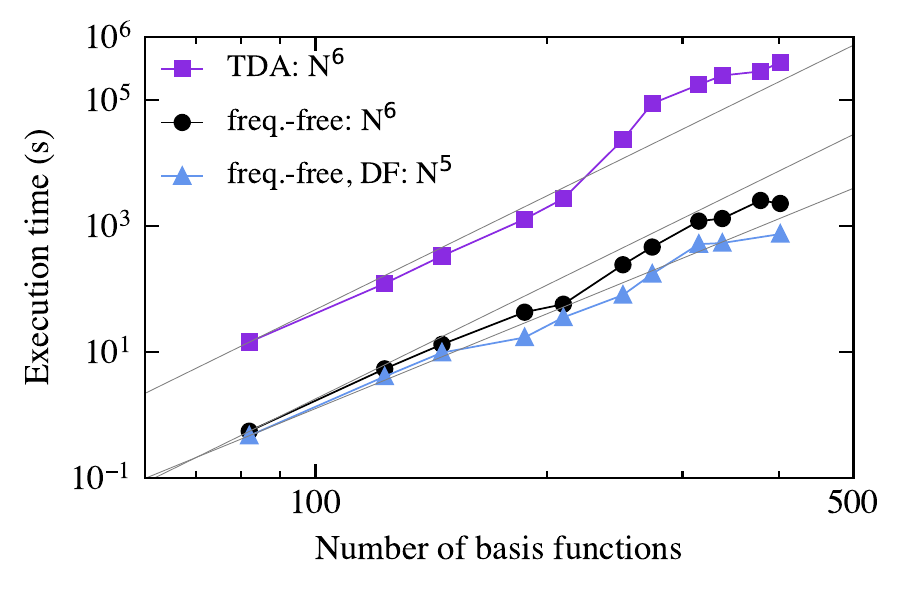}
  \caption{
Timings of the sum-over-states (``TDA''), frequency-free
(``freq.-free''), and density-fitted frequency-free (``freq.-free, DF'') implementations of the
BSE for calculating the first excitation of a series of alkenes
C$_{2n}$H$_{2n+2}$ in the aug-cc-pVDZ basis set~\cite{Dunning1989, Kendall1992, Woon1993}, up
to C$_{12}$H$_{14}$, with about 400 basis functions.
For the sum-over-states implementation, we timed the diagonalization of the TDA matrix,
which dominates the cost of GW and BSE calculations;
for the frequency-free implementations, we timed all 8-16 matrix-vector multiplications required for
convergence of the Davidson algorithm for the BSE step.
Calculations were performed on a single core of an Intel Xeon Gold 6126 2.6 GHz (Skylake) central processing unit (CPU), and density-fitted calculations used the aug-cc-pVDZ-RI auxiliary basis set.}
 \label{fig:scaling}
\end{figure}

Eigenvalues and eigenvectors of $\bm{\mathcal{H}}$ can be obtained by iterative diagonalization
using, e.g., the Davidson algorithm.
Writing the solution vector as $\bm{R} = [r_i^a, r_{ij}^{ab}, \bar{r}_{ij}^{ab}]$,
matrix-vector multiplication is given by $\bm{\mathcal{H}}\bm{R} = \bm{\sigma}$, with  
\begin{subequations}
\begin{align}
\begin{split}
\sigma_{i}^{a}
&= (E_a-E_i)r_{i}^{a}
    +\sum_{jb}\left[\kappa(ia|jb) - (ij|ab)\right] r_{j}^{b} \\
&\hspace{1em} -\sqrt{2}\sum_{dkc} ( kc|ad) r_{ik}^{dc} - \sqrt{2}\sum_{lkc}(il|ck) \bar{r}_{lk}^{ac}
\end{split}
\\
\begin{split}
\sigma_{lk}^{dc}
&= (E_d-E_l+\varepsilon_c-\varepsilon_k)r_{lk}^{dc} \\
&\hspace{1em} + \sqrt{2}\sum_{i}(il|ck)  r_{i}^{d} + 2\sum_{ia} (kc|ia) r_{li}^{da}
\end{split} 
\\
\begin{split}
\bar{\sigma}_{lk}^{dc}
&= (E_d-E_l+\varepsilon_c-\varepsilon_k)\bar{r}_{lk}^{dc} \\
&\hspace{1em} + \sqrt{2}\sum_{a}(kc|ad)  r_{l}^{a} + 2\sum_{ia} (kc|ia)\bar{r}_{li}^{da}.
\end{split} 
\end{align}
\end{subequations}
Eigenvalues found in this way naturally include dynamical screening and are exactly the same
as those from the conventional dynamically screened BSE (within the TDA).
As written, the cost of the above matrix-vector products is $O(O^3V^3)$, where $O$ and $V$ are the number
of occupied and virtual orbitals in the single-particle basis, or $O(N^6)$ generically.
While the prefactor is significantly smaller (see below), this reformulation
exhibits the same asymptotic scaling as full diagonalization of the RPA matrix and
explicit evaluation of the dynamically screened Coulomb interaction~(\ref{eq:Kd}).
However, the absence of exchange-type integrals in the direct RPA enables a scaling reduction
through the use of density fitting with $N_\mathrm{aux}$ auxiliary basis functions, 
$(pq|rs) \approx \sum_P L_{pq}^{P} L_{rs}^{P}$.  This
leads to a reformulation of the worst scaling terms, e.g.
\begin{equation}
\sigma_{lk}^{dc} = 2\sum_{iaP} L_{kc}^{P} L_{ia}^{P} r_{li}^{da} + \ldots
\end{equation}
which now has two $O(N_\mathrm{aux}O^2V^2)$ steps or $O(N^5)$ overall. 

In Fig.~\ref{fig:scaling}, we show the execution time of dynamical BSE calculations for
a series of alkenes of increasing length. As long as only a few BSE eigenvectors are required,
the reformulation to an iterative eigenvalue
problem is seen to reduce the $N^6$ prefactor by about two orders of magnitude. The use
of density fitted integrals changes the scaling to $O(N^5)$. For a system with hundreds of basis functions, 
selected BSE eigenvalues can be found in a few minutes on a single core.
Further timing details are given in the caption of Fig.~\ref{fig:scaling}.
We recall that an exact BSE calculation (within a basis) requires, as input, all GW quasiparticle
eigenvalues. As long as an $O(N^4)$ GW method is used to find the $O(N)$ GW eigenvalues, then this
initial calculation also has $O(N^5)$ scaling, and so the asymptotic scaling of the full GW/BSE
calculation is $O(N^5)$.

\begin{figure}[t]
  \centering
    \includegraphics[scale=0.85]{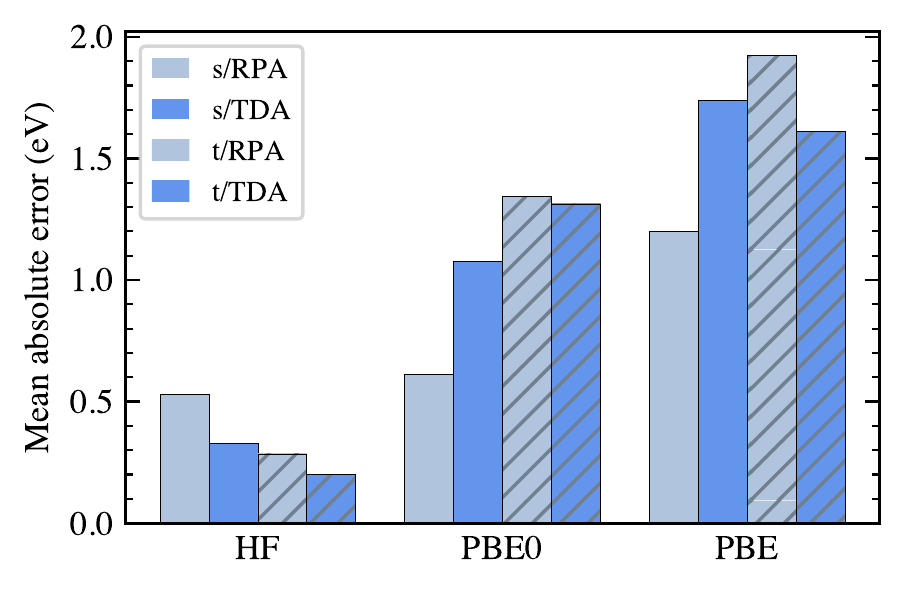}
  \caption{
Mean absolute errors in singlet (s) and triplet (t) excitation energies calculated using the
dynamical GW/BSE with RPA~\cite{LoosBlase2020} and TDA screening, starting
from three mean-field references. Statistics were collected for all excitations 
with less than 50\% double excitation character~\cite{Harbach2014} for the 25 
smallest molecules in the Thiel set~\cite{Schreiber2008} in the def2-TZVP basis~\cite{Weigend2005}.  
Error is calculated with respect to the first Thiel set theoretical best
estimates.~\cite{SilvaJunior2010}
}
  \label{fig:tda_rpa}
\end{figure}

In most GW/BSE calculations, RPA screening is used without 
the TDA (even though the TDA is commonly made in a subsequent BSE calculation),
but our frequency-free implementation is simplest with the TDA.
Therefore, we assessed the accuracy of this approximation by comparing to the 
``theoretical best estimate'' excitation energies of the molecules in the
Thiel set~\cite{Schreiber2008}. In Fig.~\ref{fig:tda_rpa}, we show that when PBE and PBE0 references are used, 
the excitation energies obtained with TDA
screening exhibit errors that are generally larger than or similar to those
obtained with RPA screening (slightly worse for singlets and slightly better for
triplets). However, a HF reference provides smaller errors
than these two DFT references. Perhaps surprisingly, 
we find that the most accurate results are those obtained
when a HF reference is combined with TDA screening, achieving an accuracy of
about 0.2-0.3 eV and empirically justifying our use of TDA screening. 
This conclusion, which is the same as we previously found for ionization potentials
of molecules within the GW approximation,~\cite{Bintrim2021} may change when GW/BSE is applied to solids,
especially metals and small gap semiconductors that are highly polarizable. 

One might ask whether the GW/BSE@HF results with TDA screening are the most accurate 
because (a) TDA screening is best for both GW and the
BSE or (b) TDA screening is best for GW but not
the BSE. To test this, we combined TDA-GW eigenvalues with an RPA-screened Coulomb
interaction in the BSE. The accuracy of the singlet excitations was not significantly different, and triplets were slightly less accurate, further
justifying our use of TDA screening for both steps of a GW/BSE@HF calculation.

As mentioned previously, the BSE with static screening (or with perturbative corrections to
account for dynamical screening~\cite{LoosBlase2020}) cannot access excited states of primarily
double excitation character. Therefore, to the best of our knowledge, the performance of GW/BSE
on double excitations has not been assessed. This is distinct from the GW approximation,
where satellite structures in the spectral function due to hole-plasmon coupling, i.e., double excitations and
higher, have been
studied extensively~\cite{Hedin1970,Guzzo2011a,Zhou2015}.
Using our frequency-free formulation of the BSE, we are well-positioned to evaluate its
performance for double excitations. As an example, we have studied the $2^1$A$_g$ excitation
of butadiene, hexatriene, and octatetraene, which is known to have substantial
double excitation character~\cite{Cave2004,Starcke2006, Romaniello2009, Angeli2011, Harbach2014, Loos2019, Manna2020}. 
Table~\ref{table:doubles} provides our GW/BSE@HF excitation energies and
the percentage contribution of doubles excitations (\%$R_2$) to the eigenvector. We also 
list theoretical best estimates (``TBE-1'')~\cite{SilvaJunior2010}
and literature values from strict and extended ADC(2), EOM-CCSD, and ADC(3)
methods. 

Overall, we see that GW/BSE@HF provides a poor description of double excitations,
overestimating their energy by about 1.5--2~eV and significantly underestimating their
double excitation character. This behavior is similar to that exhibited by strict ADC(2)
and EOM-CCSD.
We tentatively conclude that the dynamical BSE provides a qualitative but not quantitative
description of doubly excited states, but more thorough testing is required.

\begin{table}[b]
\caption{Excitation energies (eV) and percentage doubles character (``\%$R_2$'') for
the $2^1$A$_\text{g}$ excitation in three alkenes (C$_{2n}$H$_{2n+2}$)
calculated with the def2-TZVP~\cite{Weigend2005} basis set. Theoretical best estimate (TBE-1) results are from
Ref.~\onlinecite{SilvaJunior2010} and all ADC and EOM-CCSD results are from Ref.~\onlinecite{Harbach2014}.}
\label{table:doubles}
\begin{ruledtabular}
\begin{tabular}{lcccccc}
     & \multicolumn{2}{c}{butadiene} & \multicolumn{2}{c}{hexatriene} & \multicolumn{2}{c}{octatetraene} \\
     & $E$       & \%$R_2$  & $E$    & \%$R_2$  & $E$   & \%$R_2$   \\ \hline
TBE-1    & 6.55     & -     & 5.09   & -     & 4.47  & -   \\
BSE      & 8.02     & 8     & 7.09   & 9     & 6.33  & 10  \\
ADC(2)-s & 7.68     & 10    & 6.72   & 12    & 5.93  & 13  \\
ADC(2)-x & 5.12     & 59    & 4.02   & 66    & 3.30  & 70  \\
ADC(3)   & 5.77     & 68    & 4.52   & 77    & 3.73  & 80  \\
EOM-CCSD & 7.42     & 24    & 6.61   & 24    & 5.99  & 21  \\
\end{tabular}
\end{ruledtabular}
\end{table}

Before concluding, we note that our frequency-independent matrix
$\bm{\mathcal{H}}$ that is represented in a basis of single and double excitations 
is superficially similar to the one appearing in configuration interaction or the ADC, 
but it is asymmetric, has two sets of double excitations, and
requires correlated GW eigenvalues as input.  With some approximations, it can
be brought to a more familiar form,
\begin{equation}
\label{eq:ham_sym}
\bm{\tilde{\mathcal{H}}} =
\left(
\begin{array}{ccc}
\mathbf{\tilde{A}} & \mathbf{V}^{\text{e}}-\mathbf{V}^{\text{h}} \\
(\mathbf{V}^{\text{e}}-\mathbf{V}^{\text{h}})^\dagger & \mathbf{\tilde{D}}
\end{array}
\right)
\end{equation}
where
\begin{subequations}
\begin{align}
\tilde{A}_{ia,jb} &= (\varepsilon_a-\varepsilon_i)\delta_{ij}\delta_{ab}
    +\kappa (ia|jb)-(ij|ab) \\
\mathbf{\tilde{D}} &= [-\bm{\varepsilon}_{\text{occ}}] \oplus 
    \bm{\varepsilon}_{\text{vir}} \oplus \mathbf{S}.
\end{align}
\end{subequations}
Note that all orbital energies are now mean-field energies; therefore 
this formulation has the advantage of not requiring
an initial GW calculation.  Instead, this formulation has a natural self-energy-like
correction,
\begin{subequations}
\label{eq:sigma_ijab}
\begin{align}
\mathbf{\tilde{\Sigma}}(\omega) &= \mathbf{V}^{\text{e}}(\omega \mathbf{I}-\mathbf{\tilde{D}})^{-1}(\mathbf{V}^{\text{e}})^\dagger
    +\mathbf{V}^{\text{h}}(\omega \mathbf{I}-\mathbf{\tilde{D}})^{-1}(\mathbf{V}^{\text{h}})^\dagger \\
\begin{split}
\tilde{\Sigma}_{ia,jb}(\omega)
    & = \delta_{ij}\sum_{km}\frac{(ik|\rho_m)(jk|\rho_m)}{\omega-(\epsilon_a-\epsilon_k+\Omega_m)} \\
    &\hspace{1em} +\delta_{ab}\sum_{cm}\frac{(ac|\rho_m)(bc|\rho_m)}{\omega-(\epsilon_c-\epsilon_i+\Omega_m)},
\end{split}
\end{align}
\end{subequations}
with a frequency argument to be evaluated at the \textit{neutral} BSE excitation energies.

Despite its attractive features and essential GW/BSE physics, our testing
(not shown) indicates that the matrix~(\ref{eq:ham_sym}) has eigenvalues that
are about 2--3 eV below our exact dynamical BSE results. 
We attribute this discrepancy to the fact that the self-energy-like correction~(\ref{eq:sigma_ijab})
has only forward time-ordered diagrams, i.e., the particle propagator is only renormalized
by two-particle+one-hole configurations and not by two-hole+one-particle configurations and vice versa
for the hole propagator. Making this approximation in a GW calculation was observed to severely
affect the GW eigenvalues, which yielded similarly poor neutral excitation energies when used in a
subsequent BSE calculation. 

To summarize, we have shown that the conventionally frequency-dependent dynamical BSE can be
exactly reformulated into a frequency-independent eigenvalue problem in an expanded
space of single and double excitations, enabling a reduced cost implementation and a study
of doubly excited states. We anticipate that this reformulation will enable future
methodological developments to account for, for example, orbital optimization or
multiconfigurational reference wavefunctions. It would be interesting to explore
extensions of the cumulant approach, which provides an improved description
of satellite peaks in the one-particle spectral function~\cite{Aryasetiawan1996,Guzzo2011a}, to two-particle response
functions~\cite{Zhou2015,Kas2016}. Similarly, while our current work has studied double excitations in gapped molecules,
it would be interesting to perform analogous studies in bulk materials to
study biexcitons, exciton-plasmon interactions, or plasmon lifetimes~\cite{Lewis2019a}
within the GW/BSE framework.

\section*{Acknowledgements}
This work was supported in part by the National Science Foundation Graduate
Research Fellowship under Grant No.~DGE-1644869 (S.J.B.) and
by the National Science Foundation under Grant No.~CHE-1848369 (T.C.B.).  We
acknowledge computing resources from Columbia University’s Shared Research
Computing Facility project, which is supported by NIH Research Facility
Improvement Grant 1G20RR030893-01, and associated funds from the New York State
Empire State Development, Division of Science Technology and Innovation (NYSTAR)
Contract C090171, both awarded April 15, 2010.
The Flatiron Institute is a division of the Simons Foundation.

\section*{Data availability statement}
The data that support the findings of this study are available from the
corresponding author upon reasonable request.

\vspace{2em}


%

\end{document}